\documentclass[aps,showpacs,preprint,nofootinbib]{revtex4}

\begin{document}



\title{\begin{flushright}
 {\footnotesize UWO\,-TH-\,06/06}
\end{flushright}Two-dimensional metric and tetrad gravities as constrained 
second order systems}

\author{R. N. Ghalati}
\email{rnowbakh@uwo.ca}
\affiliation{Department of Applied Mathematics,
University of Western Ontario, London, N6A~5B7 Canada}

\author{N. Kiriushcheva}
\email{nkiriush@uwo.ca}
\affiliation{Department of Applied Mathematics,
University of Western Ontario, London, N6A~5B7 Canada}

\author{S. V. Kuzmin}
\email{skuzmin@uwo.ca}
\affiliation{Department of Applied Mathematics,
University of Western Ontario, London, N6A~5B7 Canada}

\date{\today}


\begin{abstract}
Using the Gitman-Lyakhovich-Tyutin generalization of the Ostrogradsky method 
for analyzing singular systems, we consider the Hamiltonian formulation of 
metric and tetrad gravities in two-dimensional Riemannian spacetime 
treating them as constrained higher-derivative theories. The algebraic 
structure of the Poisson brackets of the constraints and the 
corresponding gauge transformations are investigated in both cases.

{\it Keywords:} Hamiltonian formulation, Einstein-Hilbert action, 
metric and tetrad gravities, two dimensions, Ostrogradsky method.
\pacs{11.10.Ef}
\end{abstract}

\maketitle

A well known major difference between gravity and other field theories is that
the former includes second order time derivatives in its usual Lagrangian
formulation. In order to pass to a Hamiltonian formulation, Pirani 
{\it et al.} 
\cite{Pirani}, \cite{PiraniSS} circumvented this situation by subtracting a 
divergence term 
from the Lagrangian which includes second order time derivatives,
leaving one with first order time derivatives of the fields in 
the Lagrangian and making it suitable for the Dirac treatment of constrained 
systems \cite{Dirac}.\footnote{ Another alternative way to circumvent this 
situation is by working with first order formalisms as briefly discussed 
later.} More specifically, they split the $d$-dimensional Einstein-Hilbert 
(EH) action in the following 
way\footnote{ $\Gamma _{\sigma \mu }^\lambda $ is the Christoffel symbol, 
$g=\det{g_{\mu\nu}}$, the signature
is $\left( +,-,-,...\right) $, and the convention for defining the Riemann 
tensor is the one used in \cite{LL} or \cite{Carmeli}.}
\begin{equation}
\label{A}S_{EH}=\int d^dx\sqrt{\left( -1\right) ^{d-1}g}\,R=\int d^dx L_{\Gamma
\Gamma }-\int d^dx V^\alpha_{\,\,\,,\alpha }\,\,, 
\end{equation}
where
\begin{eqnarray*}
\label{B}L_{\Gamma \Gamma }&=&\sqrt{\left( -1\right) ^{d-1}g}\,g^{\mu \nu
}\left( \Gamma _{\sigma \mu }^\lambda \Gamma _{\lambda \nu }^\sigma -\Gamma
_{\sigma \lambda }^\lambda \Gamma _{\mu \nu }^\sigma \right) \,,\\
\label{C}V^\alpha &=&\sqrt{\left( -1\right) ^{d-1}g}\left( g^{\alpha \mu
}\Gamma _{\mu \nu }^\nu -g^{\mu \nu }\Gamma _{\mu \nu }^\alpha \right)\,, 
\end{eqnarray*}
and then neglect the contribution of the surface term 
$V^\alpha_{\,\,\,,\alpha }$ as its inclusion
has no effect on equations of motion. As was emphasized in \cite{Pirani}, 
a problem with this approach is that the part of the action 
remaining after elimination of the surface term, the $\Gamma \Gamma$-part,
is not invariant under a general coordinate transformation. This can lead to
inconsistencies in the Hamiltonian treatment of this system. 

The importance of a surface term for 
the Hamiltonian formulation of General Relativity based on the ADM slicing of 
spacetime was emphasized by Regge and Teitelboim \cite{RT} and in 
the path-integral approach by Hawking \cite{Hawking}.

In two dimensions the gravitational action, given in (\ref{A}), is a 
topological quantity. The equations of motion for $g_{\mu \nu}$ in two 
dimensions are trivial identities, putting no restriction on the metric. No 
matter how the Cauchy problem is formulated, the gravitational 
fields are arbitrary functions of spacetime; as any possible 
configuration of metric extermizes the action, the latter is a constant.

However, in two dimensions not only does the $L_{\Gamma \Gamma}$ part of 
(\ref{A}) not vanish \cite{KK2}, but it also gives rise to 
a striking but consistent Hamiltonian treatment of two dimensional gravity. 
Applying the Dirac formalism to the $L_{\Gamma \Gamma}$ leads to a gauge 
transformation \cite{KK2} which is simply that one can add any arbitrary 
tensor to the metric tensor. This is consistent with the metric tensor being 
arbitrary.

In contrast, the total two-dimensional Lagrangian of (\ref{A}), when 
expressed in terms of tetrad variables, is a total derivative. This fact 
has eroneously led to the conclusion that a Hamiltonian formulation 
of two-dimensional gravity is impossible.\footnote{ This conclusion was 
supported by writing down the Lagrangian for metric gravity in special 
coordinate systems (such as the conformal frame), where 
the $\Gamma \Gamma$-part vanishes, and then {\it wrongly} generalizing this 
to any coordinate system.} However, a Hamiltonian treatment of (\ref{A}) 
is possible in the metric formulation on account of  
the inequivalence of the metric and tetrad formulations of gravity 
\cite{Einstein1928}.\footnote{ As the action is a total derivative in the 
tetrad formulation, any possible transformation 
of tetrads is a candidate for a gauge transformation.}

One approach to the Hamiltonian formulation of gravity which avoids the 
problem of second order time derivatives is to work with the  
first order formulations of gravity, such as the metric-affine connection 
formulation of Einstein \cite{Einstein}, or the tetrad-spin connection 
formulation \cite{Schwinger} - \cite{GrumKV}. 

For the two-dimensional case, however, the equivalence of the metric-affine 
connection or tetrad-spin coonection formulation with the original second 
order formulations is obscured,
as the metric (tetrad) no longer uniquely determines the affine (spin) 
connection. This is explained in \cite{Mann} - \cite{KK1}. 
Fortunately, there is still one more approach that is applicable to 
theories with higher-order time derivatives which involves reducing the 
order of time derivatives by introducing extra fields, following Ostrogradsky 
\cite{Ostr} (see also \cite{Whittaker}). The application of such a formulation 
to two-dimensional gravity is the subject of this article.

The Hamiltonian formulation of higher-order theories was considered more
than one and a half centuries ago by Ostrogradsky \cite{Ostr} for the case of
non-singular systems, and was several times rediscovered by others.\footnote
{ The criticism that can sometimes be found on the limitation of the original
Ostrogradsky results to non-singular Lagrangians is not entirely correct
because he clearly stated this restriction in his analysis.} Before 
generalization to singular 
systems, this approach was discussed for a few higher-order theories such as 
Podolsky electrodynamics \cite{PK1944} - \cite{GP}, and
scalar fields with higher-derivative couplings \cite{BG1977}; and was first 
applied to the EH action in four dimensions by Dutt and Dresden \cite{DD}.

The first systematic generalization of the Ostrogradsky method to singular 
systems was considered by Gitman, Lyakhovich and Tyutin (GLT) \cite{GT1983} 
(see also \cite{GT}), and was applied to the Hamiltonian formulation
of generalized Yang-Mills theory \cite{GLT1985}, \cite{Galvaao}, and 
higher-derivative gravity \cite{BL1987}.

By a suitable introduction of extra fields, this generalization of the 
Ostrogradsky method allows one to reformulate a problem with higher 
derivatives in such a way that the Dirac procedure which was originally 
capable of handling only theories with first-order time derivatives 
\cite{Dirac} can be used for singular, in particular, gauge theories 
with higher order time derivatives. Details of this generalization and its 
variations appearing in the literature \cite{HO}, \cite{BGPR}, and 
\cite{Nesterenko} are beyond the scope of this 
article. We instead provide an illustration of how it can be used by 
considering the EH action in $d$ dimensions and then specializing to $d=2$ 
dimensions.

The Lagrangian of the (metric) EH action depends on second derivatives of the 
fields
\begin{equation}
\label{1}L_g=L_g\left( g_{\alpha \beta };g_{\alpha \beta ,0};g_{\alpha \beta
,k};g_{\alpha \beta ,00};g_{\alpha \beta ,0k};g_{\alpha \beta ,km}\right)\, , 
\end{equation}
where we have separated time derivatives of the fields to make subsequent 
discussion more transparent.

If, following GLT, the additional variables
\begin{equation}
\label{2}G_{\alpha \beta }=\dot g_{\alpha \beta }\,;\, v_{\alpha \beta }=\dot
G_{\alpha \beta } 
\end{equation}
are introduced, the Lagrangian (\ref{1}) can be represented in the following 
way
\begin{equation}
\label{3}L_g^v=L_g^v\left( g_{\alpha \beta };G_{\alpha \beta };g_{\alpha
\beta ,k};v_{\alpha \beta };G_{\alpha \beta ,k};g_{\alpha \beta ,km}\right)\,, 
\end{equation}
where only first-order time derivative of fields appears. As it will be shown 
below, this Lagrangian describes the same dynamical system as that of 
(\ref{1}) if it is supplemented by the conditions of (\ref{2})
\begin{equation}
\label{5}S=\int \tilde L_g^vd^dx =\int \left[ L_g^v+\pi ^{\alpha \beta
}\left( \dot g_{\alpha \beta }-G_{\alpha \beta }\right) +\Pi ^{\alpha \beta
}\left( \dot G_{\alpha \beta }-v_{\alpha \beta }\right) \right] d^dx\,. 
\end{equation}

The Lagrange multipliers  $\pi^{\alpha\beta}$ and $\Pi^{\alpha\beta}$  
act as momenta conjugate to $g_{\alpha\beta}$ and $G_{\alpha\beta}$ 
respectively. Variation of this action results in the
following equations of motion
\begin{equation}
\label{6}\frac{\delta \tilde L_g^v}{\delta \pi ^{\alpha \beta }}=\dot
g_{\alpha \beta }-G_{\alpha \beta }=0\,, 
\end{equation}
\begin{equation}
\label{7}\frac{\delta \tilde L_g^v}{\delta \Pi ^{\alpha \beta }}=\dot
G_{\alpha \beta }-v_{\alpha \beta }=0\,, 
\end{equation}
\begin{equation}
\label{8}\frac{\delta \tilde L_g^v}{\delta g_{\alpha \beta }}=\frac{\partial
L_g^v}{\partial g_{\alpha \beta }}-\partial _k\frac{\partial L_g^v}{\partial
g_{\alpha \beta ,k}}+\partial _k\partial _m\frac{\partial L_g^v}{\partial
g_{\alpha \beta ,km}}-\dot \pi ^{\alpha \beta }=0\,, 
\end{equation}
\begin{equation}
\label{9}\frac{\delta \tilde L_g^v}{\delta G_{\alpha \beta }}=\frac{\partial
L_g^v}{\partial G_{\alpha \beta }}-\partial _k\frac{\partial L_g^v}{\partial
G_{\alpha \beta ,k}}-\pi ^{\alpha \beta }-\dot \Pi ^{\alpha \beta }=0\,, 
\end{equation}
\begin{equation}
\label{10}\frac{\delta \tilde L_g^v}{\delta v_{\alpha \beta }}=\frac{%
\partial L_g^v}{\partial v_{\alpha \beta }}-\Pi ^{\alpha \beta }=0\,, 
\end{equation}
provided all variations vanish on the boundary. These are equivalent 
to the Lagrange equations following from variation of the 
original EH action (\ref{1})\footnote{ Note that there is no 
coefficient ``2'' in the fifth term of (\ref{11})
since the symmetry of double derivatives $\partial _0\partial _k$ has been 
implicitly taken into account in (\ref{1}) (for an explicit form of 
(\ref{1}) see (\ref{25}) below).}
$$
\frac{\delta L_g}{\delta g_{\alpha \beta }}=\frac{\partial L_g}{\partial
g_{\alpha \beta }}-\partial _0\frac{\partial L_g}{\partial g_{\alpha \beta
,0}}-\partial _k\frac{\partial L_g}{\partial g_{\alpha \beta ,k}}+ 
$$
\begin{equation}
\label{11}\partial _0\partial _0\frac{\partial L_g}{\partial g_{\alpha \beta
,00}}+\partial _0\partial _k\frac{\partial L_g}{\partial g_{\alpha \beta ,0k}%
}+\partial _k\partial _m\frac{\partial L_g}{\partial g_{\alpha \beta ,km}}=0 
\,. 
\end{equation}
This is easy to verify by differentiating (\ref{10}) with respect to time and 
substituting $\dot{\Pi}^{\alpha \beta}$ into (\ref{9}). We differentiate again 
with respect to time to get
\begin{equation}
\label{12}\dot \pi ^{\alpha \beta }=\partial _0\frac{\partial L_g^v}{%
\partial G_{\alpha \beta }}-\partial _0\partial _k\frac{\partial L_g^v}{%
\partial G_{\alpha \beta ,k}}-\partial _0\partial _0\frac{\partial L_g^v}{%
\partial v_{\alpha \beta }}\,. 
\end{equation}
Substituting of $\dot{\pi}^{\alpha \beta}$ into (\ref{8}) and 
supplementing the result with solutions of (\ref{6}) and (\ref{7}) finally 
establishes equivalence. The advantage of the new Lagrangian is that it 
contains only first-order time derivative of fields, permitting us to write 
the Hamiltonian in the usual way
\begin{equation}
\label{15}H=\pi ^{\alpha \beta }\dot g_{\alpha \beta }+\Pi ^{\alpha \beta
}\dot G_{\alpha \beta }-\tilde L_g^v\,, 
\end{equation}
leading to
\begin{equation}
\label{14}H=\pi ^{\alpha \beta }G_{\alpha \beta }+\Pi ^{\alpha \beta
}v_{\alpha \beta }-L_g^v\,. 
\end{equation}
This can be verified by a simple rearrangement of the terms in 
$\tilde{L}_g^v$. The fundamental Poisson brackets (PB) associated with this 
Hamiltonian are
\begin{equation}
\label{15a}\left\{ g_{\alpha \beta },\pi ^{\mu \nu }\right\} =\left\{
G_{\alpha \beta },\Pi ^{\mu \nu }\right\} =\frac 12\left( \delta _\alpha ^\mu 
\delta _\beta ^\nu +\delta _\beta ^\mu
\delta _\alpha ^\nu \right)\,, 
\end{equation}
and the PB for any two functionals $A$ and $B$ of the canonical variables is
\begin{equation}
\label{15b}\left\{ A,B\right\} =\frac{\delta A}{\delta g_{\alpha \beta }}%
\frac{\delta B}{\delta \pi ^{\alpha \beta }}+\frac{\delta A}{\delta
G_{\alpha \beta }}\frac{\delta B}{\delta \Pi ^{\alpha \beta }}-\left(
A\leftrightarrow B\right). 
\end{equation}
As an indication of the singular nature of this Lagrangian, we observe that 
the equations of motion following from the Hamiltonian formulation are
\begin{equation}
\label{16}\dot g_{\alpha \beta }=\left\{ g_{\alpha \beta },H\right\} =\frac{%
\delta H}{\delta \pi ^{\alpha \beta }}=G_{\alpha \beta }\,, 
\end{equation}
\begin{equation}
\label{17}\dot G_{\alpha \beta }=\left\{ G_{\alpha \beta },H\right\} =\frac{%
\delta H}{\delta \Pi ^{\alpha \beta }}=v_{\alpha \beta } \,,
\end{equation}
\begin{equation}
\label{18}\dot \pi ^{\alpha \beta }=\left\{ \pi ^{\alpha \beta },H\right\} =-%
\frac{\delta H}{\delta g_{\alpha \beta }}=\frac{\delta L_g^v}{\delta
g_{\alpha \beta }} \,,
\end{equation}
\begin{equation}
\label{19}\dot \Pi ^{\alpha \beta }=\left\{ \Pi ^{\alpha \beta },H\right\} =-%
\frac{\delta H}{\delta G_{\alpha \beta }}=\frac{\delta L_g^v}{\delta
G_{\alpha \beta }} \,,
\end{equation}
which are not equivalent to the equations of motion following from the 
Lagrangian formulation. This is because equation (\ref{10})
\begin{equation}
\label{20}\frac{\partial L_g^v}{\partial v_{\alpha \beta }}-\Pi ^{\alpha
\beta }=0 \,
\end{equation}
is missing. If the Lagrangian were non-singular (Hessian were not 
zero), one would be able to solve (\ref{20}) for $v_{\alpha \beta }$ and 
substitute the solution back into (\ref{17}) to get a consistent 
Hamiltonian formulation for the dynamics of the two pairs of conjugate 
variables $\left(g_{\alpha \beta },\pi ^{\alpha \beta }\right) $ and 
$\left( G_{\alpha \beta},\Pi ^{\alpha \beta }\right) $, without any of 
constraints. However, for the EH Lagrangian, because it is linear in the 
second-order derivatives, Eq. (\ref{20}) cannot be solved. 
Therefore Eq. (\ref{20}) has to be supplemented to the 
Hamiltonian formulation as a set of primary constraints.

We now demonstrate this explicitly, starting in $d$ dimensions, and then 
switching to two dimensions, which is the main concern of this article. 
If the Riemann 
tensor is written explicitly in terms of the metric tensor and its 
derivatives \cite{DD}, 
the EH Lagrangian splits into two parts in the following way 
\begin{equation}
\label{21}L=\sqrt{(-1)^{d-1}g}\,R=A^{\alpha \beta \mu \nu }g_{\alpha \beta ,
\mu \nu }+B^{\alpha \beta \mu \nu \gamma \rho }g_{\alpha \beta ,\gamma }
g_{\mu \nu,\rho }\,, 
\end{equation}
where
\begin{equation}
\label{22}A^{\alpha \beta \mu \nu }=\sqrt{(-1)^{d-1}g}\left( g^{\alpha \mu }
g^{\beta \nu }-g^{\alpha \beta }g^{\mu \nu }\right) 
\end{equation}
and
$$
B^{\alpha \beta \mu \nu \gamma \rho }=-\frac 14\sqrt{(-1)^{d-1}g}\left( 
g^{\alpha
\beta }g^{\mu \nu }g^{\gamma \rho }-3g^{\alpha \mu }g^{\beta \nu }g^{\gamma
\rho }\right. 
$$
\begin{equation}
\label{23}\left. +2g^{\alpha \rho }g^{\beta \nu }g^{\gamma \mu }+4g^{\alpha
\gamma }g^{\mu \rho }g^{\beta \nu }-4g^{\alpha \gamma }g^{\beta \rho }g^{\mu
\nu }\right)\,. 
\end{equation}
After introducing the additional variables of (\ref{2}), 
the Lagrangian becomes
\begin{equation}
\label{24}L=\pi ^{\alpha \beta }\dot g_{\alpha \beta }+\Pi ^{\alpha \beta
}\dot G_{\alpha \beta }-H \,,
\end{equation}
with
$$
H=\pi ^{\alpha \beta }G_{\alpha \beta }+v_{\alpha \beta }\left( \Pi ^{\alpha
\beta }-A^{\alpha \beta 00}\right) -\left( A^{\alpha \beta 0k}+A^{\alpha
\beta k0}\right) G_{\alpha \beta ,k}-A^{\alpha \beta km}g_{\alpha \beta ,km} 
$$
\begin{equation}
\label{25}-B^{\alpha \beta \mu \nu 00}G_{\alpha \beta }G_{\mu \nu }-\left(
B^{\alpha \beta \mu \nu 0k}+B^{\mu \nu \alpha \beta k0}\right) G_{\alpha
\beta }g_{\mu \nu ,k}-B^{\alpha \beta \mu \nu km}g_{\alpha \beta ,k}g_{\mu
\nu ,m} \,. 
\end{equation}
Here it can be explicitly seen how the second term in this expression 
leads to the aforementioned  primary constraints of (\ref{10}) 
\begin{equation}
\label{27}P^{\alpha \beta } \equiv \Pi ^{\alpha \beta }-A^{\alpha \beta 00}\,, 
\end{equation}
where $A^{\alpha\beta\mu\nu}$ is given by (\ref{22}). At this point
we set $d=2$. By  making use of the fact that in two dimensions
$$
g^{11}g^{00}-g^{01}g^{01}=1/g \,,$$ one obtains the following constraints 
from (\ref{27})
\begin{equation}
\label{28}P^{11}=\Pi ^{11}-\frac 1{\sqrt{-g}}\approx
0\,, P^{00}=\Pi ^{00}\approx 0\,, P^{01}=\Pi ^{01}\approx 0\,. 
\end{equation}
These primary constraints $P^{\alpha \beta}$ obviously constitute a first 
class system with a simple algebra of PBs
\begin{equation}
\label{34}\left\{ P^{\alpha \beta},P^{\mu \nu}\right\} =0. 
\end{equation}
The secondary constraints are determined by requiring that the 
primary constraints be preserved in time
\begin{equation}
\label{35}\dot P^{\alpha \beta }=\left\{ P^{\alpha \beta },H\right\}
\equiv S^{\alpha \beta } = 0 \,,
\end{equation}
leading to the following secondary constraints
\begin{equation}
\label{30}S^{00}=-\pi ^{00}-\frac 12\frac{g^{00}}{\sqrt{-g}}G_{11}+\frac 12%
\frac{g^{01}}{\sqrt{-g}}g_{11,1}+\frac{g^{00}}{\sqrt{-g}}g_{01,1}\,, 
\end{equation}
\begin{equation}
\label{31}S^{01}=-\pi ^{01}-\frac 12\frac{g^{01}}{\sqrt{-g}}G_{11}-\frac 12%
\frac{g^{00}}{\sqrt{-g}}g_{00,1}\,, 
\end{equation}
\begin{equation}
\label{32}S^{11}=-\pi ^{11}-\frac 12\frac{g^{11}}{\sqrt{-g}}G_{11}-\frac 12%
\frac{g^{01}}{\sqrt{-g}}g_{00,1} \,.
\end{equation}
It is not difficult to show that these new constraints (\ref{30}-\ref{32}) 
commute with the set of primary constraints
\begin{equation}
\label{36}\left\{ P^{\alpha \beta},S^{\mu \nu}\right\} =0\,, 
\end{equation}
and the PBs among secondary constraints $S^{\alpha \beta}$ also vanish
\begin{equation}
\label{37}\left\{ S^{\alpha \beta},S^{\mu \nu}\right\} =0. 
\end{equation}
Thus all primary and secondary constraints are first class.
Using (\ref{30}-\ref{32}) one can write the Hamiltonian (\ref{25}) in the 
following way
\begin{equation}
\label{26}H=v_{\alpha \beta }P^{\alpha \beta }-G_{\alpha \beta
}S^{\alpha \beta }+\tilde H \,,
\end{equation}
with $\tilde H$ including only spatial derivatives
$$
\tilde H=-A^{\alpha \beta 11}g_{\alpha \beta ,11}-B^{\alpha \beta \mu \nu
11}g_{\alpha \beta ,1}g_{\mu \nu ,1} 
$$
\begin{equation}
\label{33}-\left[ \left( A^{\alpha \beta 01}+A^{\alpha \beta 10}\right)
G_{\alpha \beta }\right] _{,1}=-\left( \frac{g_{00,1}-2G_{01}}{\sqrt{-g}}%
\right) _{,1} \,.
\end{equation}

The next step is to consider the time derivative of the secondary constraints 
to see if any new constraints arise
\begin{equation}
\label{38}\dot S^{\alpha \beta }=\left\{ S^{\alpha \beta },H\right\}. 
\end{equation}
Using the form of the Hamiltonian in (\ref{26}) and the fact that 
primary and secondary constraints are all first class, these PBs are equal 
to $\left\{ \pi ^{\alpha \beta },\tilde H\right\}$. 
These latter brackets can be obtained using (\ref{15}) and assuming that 
the fields and their spatial derivatives vanish rapidly at infinity. An 
alternative is to treat $\tilde H$ as a spatial surface term and totally 
neglect it. Either method shows that these brackets vanish, leading to
the closure of the Dirac procedure with six
first class constraints with zero PBs among them. As there are six independent 
fields $g_{\alpha \beta }$ and $%
G_{\alpha \beta }$, and there are six first class constraints, there is  
zero degrees of freedom. Using the Castellani procedure \cite{Cast},
one can restore the gauge transformation of fields by building the gauge 
generator $G\left( \varepsilon \right) $ corresponding to the complete 
set of first class constraints. In much the same way as in \cite{KKM1},
\cite{KKM2}, and 
\cite{KK1}, we find that the gauge generator $G\left( \varepsilon \right)$ is 
\begin{equation}
\label{39}G\left( \varepsilon \right) =\int dx \left( -\varepsilon _{\alpha 
\beta }\left\{ P^{\alpha \beta },H\right\}  +\dot \varepsilon _{\alpha
\beta }P^{\alpha \beta }\right).
\end{equation}
This gives the following gauge transformations with arbitrary gauge parameters 
$\epsilon_{\alpha\beta}$
\begin{equation}
\label{40}\delta g_{\alpha \beta }=\left\{ g_{\alpha \beta },G\left(
\varepsilon \right) \right\} =\varepsilon _{\alpha \beta } 
\end{equation}
and 
\begin{equation}
\label{40-b}\delta G_{\alpha \beta }=\left\{ G_{\alpha \beta },G\left(
\varepsilon \right) \right\} =\dot{\varepsilon} _{\alpha \beta }\,. 
\end{equation}
This is consistent with (\ref{2}). The transformations of (\ref{40}) 
are exactly the same as the transformations obtained in \cite{KK2}, where 
only the $L_{\Gamma \Gamma}$ part of the EH Lagrangian was considered. 
However, there is a difference in the number of fields and the structure of 
the constraints 
in these two methods.\footnote{ In passing, we note that there are 
a few possible variants of the generalized Ostrogradsky method that lead to 
the same transformations. For example, it is enough to introduce extra fields 
only for those $g_{\alpha \beta }$ components that have second-order time 
derivatives (only $g_{11}$ in two dimensions). In this case, the analysis 
leads to four first class constraints (three primary and one secondary) for 
the four fields which again leads to zero degrees of freedom and the same 
gauge tramsformation as (\ref{40}).}

We now investigate the application of the Ostrogradsky 
method to tetrad gravity. This merits interest since, 
as has already been discussed, the Lagrangian for tetrad gravity is a pure 
surface term in two dimensions, making its treatment using
the method employed by Pirani et al \cite{Pirani}, \cite{PiraniSS} impossible. 

If one substitutes into the EH Lagrangian (\ref{21}) the expression for 
$g_{\mu\nu}$ in terms of tetrads
\begin{equation}
\label{41}g_{\mu \nu }=e_\mu ^ae_\nu ^b\eta _{ab}\,, 
\end{equation} 
one obtains the second order Lagrangian for gravity in terms of tetrads,
which can be analyzed using the Ostrogradsky method.\footnote{ $\eta _{ab}$ 
- Minkowsky 
metric, $\mu=0,...,d-1$ are world indices and $a=0,...d-1$ are tetrad 
indices.} The result of this substitution can be written in  
the compact form \cite{KK3}
\begin{equation}
\label{51}L_e=\left( 2\varepsilon ^{ab}\varepsilon ^{\nu \rho }e_a^\mu
e_{b\rho ,\nu }\right) _{,\mu }=2\varepsilon ^{ab}\varepsilon ^{\nu \rho
}\left( e_a^\mu e_{b\rho ,\nu \mu }-e_a^\sigma e_c^\mu e_{\sigma ,\mu
}^ce_{b\rho ,\nu }\right)\,, 
\end{equation}
where $\varepsilon$ is the totally antisymmetric tensor $\varepsilon
^{01}=\varepsilon ^{\left( 0\right) \left( 1\right) }=1$.\footnote{ $\left( 
{}\right) $ brackets distinguish explicit values of tetrad indices.} 
Introducing, as in the metric case, additional variables
\begin{equation}
\label{52}E_\mu ^a=\dot e_\mu ^a;\,\,v_\mu ^a=\dot E_\mu ^a 
\end{equation}
and performing manipulations similar to those used in obtaining (\ref{15}),
we get
\begin{equation}
\label{53}\tilde L_e^v=\pi _a^\mu \dot e_\mu ^a+\Pi _a^\mu \dot E_\mu ^a-H \,,
\end{equation}
with
\begin{equation}
\label{54}H=\pi _a^\mu E_\mu ^a+\Pi _a^\mu v_\mu ^a-L_e^v. 
\end{equation}
We define the following fundamental PBs 
\begin{equation}
\label{57}\left\{ e_\mu ^a,\pi _b^\nu \right\} =\left\{ E_\mu ^a,\Pi _b^\nu
\right\} =\delta _b^a\delta _\mu ^\nu \,
\end{equation}
and have
$$
L_e^v=2\varepsilon ^{ab}e_a^0v_{b1}+2\varepsilon
^{ab}e_a^1E_{b1,1}-2\varepsilon ^{ab}e_a^0E_{b0,1}-2\varepsilon
^{ab}e_a^1e_{b0,11} 
$$
\begin{equation}
\label{55}-2\varepsilon ^{ab}e_a^\sigma e_c^0E_\sigma ^cE_{b1}-2\varepsilon
^{ab}e_a^\sigma e_c^1e_{\sigma ,1}^cE_{b1}+2\varepsilon ^{ab}e_a^\sigma
e_c^0E_\sigma ^ce_{b0,1}+2\varepsilon ^{ab}e_a^\sigma e_c^1e_{\sigma
,1}^ce_{b0,1} \,.
\end{equation}
After rearrangement of the terms in (\ref{55}), one obtains 
$$
H=\Pi _a^\mu v_\mu ^a-2\varepsilon ^{ab}e_a^0v_{b1}+\pi _a^\mu E_\mu
^a+2\varepsilon ^{ab}e_a^\sigma e_c^0E_\sigma ^cE_{b1}-2\varepsilon
^{ab}e_a^\sigma e_c^0e_{b0,1}E_\sigma ^c 
$$
\begin{equation}
\label{55A}+2\varepsilon ^{ab}e_a^\nu e_c^0e_{\nu ,1}^cE_{b0}-\left[
2\varepsilon ^{ab}\left( e_a^1e_{b0,1}+e_a^1E_{b1}-e_a^0E_{b0}\right)
\right] _{,1} \,.
\end{equation}
The first two terms in this expression give rise to a set of primary 
constraints $ P_a^\mu $
\begin{equation}
\label{56}P_{\left( 0\right) }^0=\Pi _{\left( 0\right) }^0\,, P_{\left(
1\right) }^0=\Pi _{\left( 1\right) }^0\,, P_{\left( 0\right) }^1=\Pi _{\left(
0\right) }^1+2e_{\left( 1\right) }^0\,, P_{\left( 1\right) }^1=\Pi _{\left(
1\right) }^1+2e_{\left( 0\right) }^0\,. 
\end{equation}
The PBs among these four primary constraints $P_a^\mu $ are
obviously zero. Conservation of primary constraints in time leads to the 
secondary constraints $S_a^\mu $
\begin{equation}
\label{58}\dot P_a^\mu =\left\{ P_a^\mu ,H\right\} \equiv S_a^\mu = 0\,.
\end{equation}
They are given by 
$$
S_{\left( 0\right) }^0=-\pi _{\left( 0\right) }^0+2e_{\left( 0\right)
}^0e_{\left( 0\right) }^0E_1^{\left( 1\right) }+2e_{\left( 1\right)
}^0e_{\left( 0\right) }^0E_1^{\left( 0\right) } 
$$
$$
+2e_{\left( 1\right) }^1e_{\left( 0\right) }^0e_{1,1}^{\left( 0\right)
}+2e_{\left( 1\right) }^1e_{\left( 1\right) }^0e_{1,1}^{\left( 1\right)
}+2\left( e_{\left( 1\right) }^0e_{\left( 1\right) }^0-e_{\left( 0\right)
}^0e_{\left( 0\right) }^0\right) e_{0,1}^{\left( 1\right) }\,, 
$$
$$
S_{\left( 1\right) }^0=-\pi _{\left( 1\right) }^0+2e_{\left( 0\right)
}^0e_{\left( 1\right) }^0E_1^{\left( 1\right) }+2e_{\left( 1\right)
}^0e_{\left( 0\right) }^0E_1^{\left( 0\right) } 
$$
$$
+2e_{\left( 0\right) }^1e_{\left( 0\right) }^0e_{1,1}^{\left( 0\right)
}+2e_{\left( 0\right) }^1e_{\left( 1\right) }^0e_{1,1}^{\left( 1\right)
}+2\left( e_{\left( 0\right) }^0e_{\left( 0\right) }^0-e_{\left( 1\right)
}^0e_{\left( 1\right) }^0\right) e_{0,1}^{\left( 0\right) }\,, 
$$
\begin{equation}
\label{60}S_a^1=-\pi _a^1+2e_{\left( 0\right) }^1e_a^0E_1^{\left( 1\right)
}+2e_{\left( 1\right) }^1e_a^0E_1^{\left( 0\right) }-2e_{\left( 0\right)
}^1e_a^0e_{0,1}^{\left( 1\right) }-2e_{\left( 1\right)
}^1e_a^0e_{0,1}^{\left( 0\right) } \,.
\end{equation}
It is fairly easy to demonstrate that
\begin{equation} 
\label{61a}\left\{ S_a^{\mu},P_b^{\nu} \right\} =0 
\end{equation}
for any pair of primary and secondary constraints, but  calculation of PBs 
among secondary constraints is slightly more involved, and requires the 
use of test functions, leading to
\begin{equation}
\label{61}\left\{ S_a^{\mu},S_b^{\nu}\right\} =0\,. 
\end{equation}
The Hamiltonian (\ref{55A}) can be expressed in terms of a linear combination 
of constraints plus a spatial derivative term
\begin{equation}
\label{62}H=v_\mu ^aP_a^\mu -E_\mu ^aS_a^\mu -\left[ 2\varepsilon
^{ab}\left( e_a^1e_{b0,1}+e_a^1E_{b1}-e_a^0E_{b0}\right) \right] _{,1}\,. 
\end{equation}
Again, there are no tertiary constraints because 
$\dot S_a^{\mu}=\left\{ S_a^{\mu},H\right\}=0$ due to eqs. (\ref{61a}) 
and (\ref{61}). The Dirac procedure is closed, with eight first class 
constraints leading to zero degrees of freedom.
Using the Castellani procedure \cite{Cast}, the generator of gauge 
transformation is
\begin{equation}
\label{63}G_e\left( \varepsilon \right) =\int dx \left( -\varepsilon _\mu ^a\
\left\{P_a^\mu ,H\right\} +\dot \varepsilon _\mu ^aP_a^\mu \right),  
\end{equation}
where $\varepsilon_\mu ^a$ are gauge parameters. This generator leads to the 
gauge transformation
\begin{equation}
\label{64}\delta _ee_\mu ^a=\left\{ e_\mu ^a,G_e\left( \varepsilon \right)
\right\} =\varepsilon _\mu ^a\,, 
\end{equation}
for the tetrad fields.

As discussed in the introduction, the transformation (\ref{64}),
as well as the transformation found for the metric formulation 
of the EH action (\ref{40}), is consistent with triviality of the equations of 
motion for the EH action in two dimensions. The number of gauge parameters for 
the metric case is three, and for the tetrad case four, the same as the number 
of independent fields in each case, leading to zero number of degrees of 
freedom according to the standard way of counting the degrees of freedom.
 
As has been discussed elsewhere, the tetrad representation of a 
spacetime leads uniquely to the metric representation \cite{Einstein1928}. 
However, the 
converse is not always true as it is not possible to uniquely determine a 
tetrad representation from a metric representation. For the same reason, 
a gauge transformation in the tetrad formulation corresponds to a unique gauge 
transformation in the metric formulation 
\begin{equation}
\epsilon_{\mu\nu}=\label{65}\delta _eg_{\mu \nu }=\delta _e\left( \eta _{ab}
e_\mu ^ae_\nu
^b\right) =\eta _{ab}e_\mu ^a\delta _ee_\nu ^b+\eta _{ab}e_\nu ^b\delta
_ee_\mu ^a=e_{b\mu }\varepsilon _\nu ^b+e_{b\nu }\varepsilon _\mu ^b\,, 
\end{equation}
while a gauge transformation in the metric formulation does not lead to  
a unique gauge transformation in the tetrad formalism even if a unique tetrad 
system is specified. This is because the three equations of (\ref{65}) 
can not be solved for the four unknowns $\varepsilon _\nu^a$ in terms of 
$\epsilon_{\mu\nu}$. In 2D the Lagrangian for tetrad gravity is a pure surface 
term but its 
Hamiltonian treatment using GLT generalization is possible. This can 
provide an alternative approach to demonstrate unequivalence of 3D tetrad
gravity and the Chern-Simons theory \cite{Matschull} (which differ by a total 
derivative) based on their Hamiltonian formulations.

To conclude, the Hamiltonian formulation of the two-dimensional EH action as a 
higher-derivative theory leads to a consistent structure of constraints, and a 
vanishing number of degrees of freedom. The gauge transformations 
of (\ref{40}) and (\ref{64})
are different from linearized coordinate transformations.\footnote{ The 
linearized coordinate transformation is $\delta g_{\mu \nu}
=-g_{\mu \lambda} \xi^\lambda_{,\nu} -g_{\nu \lambda} \xi^\lambda_{,\mu} -
\xi^\lambda g_{\mu \nu , \lambda}$ \,\,\cite{LL}, and the corresponding 
algebra of PB among constraints is known as Dirac, hypersurface-deformation, 
or diffeomorphism algebra. The last name actually is abuse of mathematical 
language because this algebra corresponds to the linearized coordinate 
transformation and not to the general one, which is called a diffeomorphism.} 
The number of constraints required to  
reproduce linearized coordinate transformations as
{\it gauge} transformation of the two-dimensional EH action, results in there 
being a negative number of degrees of freedom \cite{Martinec} which is clearly 
unacceptable. Obtaining diffeomorphism invariance as a gauge transformation 
leads to discrepancies in the number of degrees of freedom and 
the number of first class constraints appearing in the Hamiltonian analysis of 
other two-dimensional models. As an example, a 
Hamiltonian analysis of a scalar field in curved spacetime $\sqrt{-g} 
g^{\mu \nu} \partial_\mu \phi \partial_\nu \phi $ 
gives five first class constraints for the four
fields $\left(g_{\mu \nu},\phi\right)$ as reported in \cite{BNeto}, 
leading to minus one number of degrees of freedom if the diffeomorphism and 
Weyl invariances are to be {\it gauge} symmetries.
However, as will be reported elsewhere, the 
treatment of this model using the Dirac procedure removes the contradiction 
arising from having a negative number of degrees of freedom and leads to a 
gauge transformation distinct from {\it the linearized coordinate 
transformation}.

In this paper the method of GLT was emlpoyed in which only the order 
of temporal derivarives was decreased (the same as was done by Dutt and 
Dresden). An alternative way is to introduce extra fields also for spatial 
derivatives of fields, an approach that may be called a covariant 
Ostrogradsky method. Recently, the lowering of order of derivatives in 
covariant form was used for three dimensional topologically massive 
electrodynamics (TME) \cite{KK} to construct its first 
order formulation. This covariant Ostrogradsky method leads to consistent 
results, despite of the conclusion of \cite{Deser}, 
and, in particular, preserves the gauge invariance which was 
explicitly demonstrated using the Dirac formalism in \cite{GKKM}. 
\\
\\
{\bf Acknowledgments}

The authors are grateful to D. G. C. McKeon for discussions and reading the 
manuscript.

\end{document}